\documentstyle[12pt,epsf]{article}

\begin{document}
\newcommand{\ket}[1] {\mbox{$ \vert #1 \rangle $}}
\newcommand{\bra}[1] {\mbox{$ \langle #1 \vert $}}
\newcommand{\bkn}[1] {\mbox{$ < #1 > $}}
\newcommand{\bk}[1] {\mbox{$ \langle #1 \rangle $}}
\newcommand{\scal}[2]{\mbox{$ \langle  #1 \vert #2 \rangle   $}}
\newcommand{\expect}[3] {\mbox{$ \bra{#1} #2 \ket{#3} $}}
\newcommand{\ki}{\mbox{$ \ket{\psi_i} $}}
\newcommand{\bi}{\mbox{$ \bra{\psi_i} $}}
\newcommand{\p} \prime
\newcommand{\e} \epsilon
\newcommand{\la} \lambda
\newcommand{\om} \omega   \newcommand{\Om} \Omega
\newcommand{\cc}{\mbox{$\cal C $}}
\newcommand{\w} {\hbox{ weak }}
\newcommand{\al} \alpha
\newcommand{\bt} \beta

\newcommand{\be} {\begin{equation}}
\newcommand{\ee} {\end{equation}}
\newcommand{\ba} {\begin{eqnarray}}
\newcommand{\ea} {\end{eqnarray}}

\def\lrD{\mathrel{{\cal D}\kern-1.em\raise1.75ex\hbox{$\leftrightarrow$}}}
\def\lr #1{\mathrel{#1\kern-1.25em\raise1.75ex\hbox{$\leftrightarrow$}}}

\begin{flushright}
LPTENS 96/10 \\
TAU 2326-96\\
October 1996\\
\end{flushright}
\vskip .8cm
\centerline{\bf THE SCHWINGER MECHANISM, THE UNRUH EFFECT}
\centerline{\bf AND THE PRODUCTION OF ACCELERATED BLACK HOLES}
\vskip 1.5cm
\centerline{R. Parentani}
\centerline{Laboratoire de Physique Th\'eorique de l'\' Ecole
Normale Sup\'erieure\footnote{Unit\'e propre de recherche du C.N.R.S. 
associee \` a  l'\' Ecole Normale Sup\'erieure et \` a l'Universit\' e de Paris Sud.},}
\centerline{24 rue Lhomond,
75.231 Paris CEDEX 05, France}
\centerline{e-mail: parenta@physique.ens.fr}
\vskip 5 truemm
\centerline{S. Massar
}
\centerline{ Raymond and Beverly Sackler Faculty of Exact Sciences,}
\centerline{ School of Physics and Astronomy,}
\centerline{Tel-Aviv University, Tel-Aviv 69978, Israel}
\centerline{e-mail: massar@ccsg.tau.ac.il}

\vskip 3 truecm
\vskip 1 truecm
{\bf Abstract }

\noindent
We compute the corrections to the transition amplitudes
of an accelerated Unruh ``box'' that arise when the accelerated box
is replaced by a ``two level ion'' immersed in a constant electric field
and treated in second quantization.
There are two kinds of corrections, those due to recoil effects induced by the
momentum transfers and those due to pair creation. 
Taken together, these corrections show that there is a direct relationship
between pair creation amplitudes described by the 
Heisenberg-Euler-Schwinger mechanism and the Unruh effect, 
i.e. the thermalisation of accelerated systems at temperature 
$a/ 2 \pi$ where $a$ is the acceleration.
In particular, there is a thermodynamical consistency
between both effects whose origin is that the euclidean action governing
pair creation rates acts {\it as} an entropy in delivering the Unruh temperature.
Upon considering pair creation of charged black holes
 in an electric field, these relationships explain why black holes are 
created from vacuum in thermal equilibrium, i.e. with their Hawking temperature 
equal to their Unruh temperature.

\newpage
 
\section{Introduction}

Quantum field theory predicts two remarkable phenomena when 
charged matter is
accelerated in a uniform electric field.

The first is
the Heisenberg-Euler-Schwinger
 mechanism\cite{schw1}\cite{schw}. That is, vacuum instability due
to spontaneous creation of charged pairs. 
When the electric field is turned on in a 
volume $V$ during a period $T$, the norm of
the overlap between the initial and the final vacua, which gives the probability
not to produce pairs, decreases as
\ba
\vert \scal{0,  out}{\ 0, in} \vert^2 = e^{-  \Gamma T} \quad {\mbox{ where} } 
\quad
 \Gamma = V ({E \over 2 \pi})^{d/2} \ln ( 1 + e^{- \pi M^2/E})
\label{01}
\ea
where $d$ is the number of space time dimensions, 
$M$ is the  mass of the scalar charged particle, $E=eE_0$ 
is the product of the charge $e$ of the particle by the 
electric field $E_0$  
(we put $\hbar = c =1$).
To allow the comparison with the second effect we shall
discuss below, it is appropriate
to notice that the pairs produced by the Schwinger mechanism 
possess a well 
defined spectrum.
Indeed, the mean number
of pairs, $N(k_{\perp})$, characterized by a transverse (with respect to the 
direction of the acceleration) momentum $k_{\perp}$ 
is given by 
\be
N(k_\perp) = C e^{- \pi (M^2 + k^2_{\perp})/E}
\label{N}
\ee
where the overall factor $C$ takes into account the phase space factor
arising from quantization in a volume $V$.
For later convenience we reexpress eq. (\ref{N}) in such a way that the 
constant $C$ cancels
\be
{ N(k_{\perp}) \over N(k_{\perp} = 0)} = e^{- \pi k_{\perp}^2 /E} = {
P(k_{\perp}) \over P(k_{\perp} = 0)} 
\label{kperp}
\ee
where we have introduced 
the probability $P(k_{\perp})$ 
that a detected particle produced from vacuum possesses that momentum.
This latter concept is more intrinsic
since it does not depend on global characteristics such as 
$V$ and $T$ nor on the rate of pair production.

The second effect concerns the physics which arises when 
(these) accelerated particles are coupled to quantum 
radiation, i.e. photons, or
more generally to some massless field $\phi$. 
Through this coupling, massive 
particles behave like detectors.  By detector we mean a 
quantum system which has two (or more) energy
levels separated by an energy gap $\Delta m$ and which can make 
transitions by emitting or absorbing a photon $\phi$. 
Therefore, these detectors probe the state of the  $\phi$ field. 

It is on that basis that Unruh\cite{Unr} proved that when such a 
detector is uniformly accelerated, it perceives the vacuum state
of the  $\phi$ field to be thermally populated 
with the temperature
\ba
T_U = a/2\pi
\label{B03}
\ea
where $a$ is the acceleration of the detector.
Such an accelerated detector will eventually reach equilibrium
with the heat bath, whereupon its two energy levels are populated with the
Boltzmanian ratio
\ba
{P_+ \over P_-}=e^{-2\pi \Delta m /a} = 
{R_{- \to +} \over R_{+ \to -}}
\label{B04}
\ea
where $P_+ (P_-)$ is the probability to be  found in the
excited (ground) state. We have also introduced the
rates of transitions $ R_{\pm \to \mp}$ 
which determine dynamically this equilibrium. 

At first sight, besides the fact that charged particles
propagate with uniform acceleration
\ba
a =e E_0/M = E/M
\label{B02}
\ea
once they are produced 
 in the constant electric field,
 the Schwinger and Unruh effects seem 
hardly related phenomena. Indeed the Schwinger effect requires a second
quantized framework for the massive charged field $\psi_{M}$ only and
 makes no references to quantized massless fields.
On the contrary, in 
 the Unruh effect, the propagation of the 
detector is described by a given classical trajectory
(this is of course an approximation, see below)
and only its internal transitions 
accompanied
by the emission or absorption of quanta of 
the radiation field
are treated quantum mechanically.
Moreover, the Unruh effect can even
be understood without introducing the detector at all.
Indeed, it suffices to reexpress the
vacuum state of the field $\phi$, i.e. Minkowski vacuum,
 in terms of its ``Rindler'' particle content.
By  Rindler quanta, one designates the quanta of the 
radiation field associated to the eigenmodes of the boost operator
which vanish beyond the horizon defined by the 
accelerated trajectory of the system\cite{Fulling}\cite{Unr}. 
Fulling found that Minkowski vacuum is a thermal distribution of
Rindler quanta. Then, Unruh proved that accelerate detectors react to Rindler 
quanta as inertial detectors react to inertial (Minkowski) 
quanta. Therefore, accelerated systems find themselves
in a thermal bath.

However, there are a number of questions 
which cannot be answered by
this kinematical analysis based on Rindler modes only. 
To reveal the aspects inevitably missed by this analysis
and to prove the necessity of considering more dynamical frameworks,
we shall proceed in three steps by posing and answering questions.

a) What is the energy balance of the accelerated-thermal 
equilibrium as seen by an inertial observer?
In order to answer this question, it is mandatory to 
abandon the description in terms of the 
Rindler modes and to use instead the Minkowski modes
of the radiation field.
The main result is that, in spite of the
equilibrium, Minkowski quanta are produced\cite{Unru2}\cite{AM} 
and their total number 
equals\cite{MaPa} the number of internal transitions of the detector.

b) Where does this energy come from?
or more locally, What is the incidence of the energy-momentum transfer 
occuring when one emission process takes place?
To answer these questions 
requires an enlargement of the dynamics. One must indeed
quantize the center of mass position of the 
detector - and thus attribute it a finite mass $M$-
 and introduce an external force such 
that the detector
 accelerates uniformly in the absence of transitions.
This is precisely the role of a constant electric field.
The main results of this enlarged dynamical framework
 are the following\cite{rec}:
 1) The transition probabilities between excited and ground state 
still satisfy eq. (\ref{B04}) when $\Delta m <\!\!< M$.
2) Due to recoil effects,
the energy flux emitted by the detector becomes rapidly
incoherent and positive. Moreover it 
is accompanied by a constant drift from uniformly
accelerated trajectories which expresses the dissipation of 
potential electric energy into radiation.

c) What are the consequences of ``second quantizing the detector'', 
i.e. of taking into account 
amplitudes
of producing pairs
of charged detectors in the electric field?  
Indeed a {\it complete} description of a quantum relativistic system 
in an external field
can only be obtained by working in a second quantized framework
(the answers delivered in point b. were based on an approximate
first quantized treatment (WKB) in which corrections in $e^{-\pi M^2/ E}$
were neglected\cite{rec}).
This further enlargement of the
dynamics {\it implies} that both the Schwinger and the Unruh effect
are encompassed in the
same model.
To analyze the consequences of this enlargement
 is the central problem addressed in the present article.

At this point, it is appropriate to consider
the emission rates of photons 
by bremsstrahlung
from an electron accelerated in a constant electric
field. 
These emission rates were analyzed
by Nikishov\cite{Niki2}\cite{Niki} a few years before Unruh's seminal work.
It is very interesting to notice that the point of view adopted by Nikishov 
was to consider these emission processes 
as describing corrections to the
Schwinger effect rather than corrections
 to the Unruh effect induced by  ``second quantizing the detector''.
This dual point of view clearly illustrates the entangled nature of 
both processes when studied in the enlarged framework. 
Nikishov showed
that the ratio of the transition rates for an electron to jump from a state
with transverse momentum $k_{\perp}$ to a state
with zero momentum accompanied by the emission of a photon 
to the inverse transition satisfies
\ba
{R_{k_{\perp } \rightarrow 0 } \over
R_{0 \rightarrow k_{ \perp } }} = e^{-{\pi  }{ k_{\perp }^2 / E} }
= e^{-{2\pi } { k_{\perp}^2 \over 2m_e}/ a_e}
\label{B05}
\ea
where $a_e=E/m_e$. 
In the second equality
we have written $ k_{\perp }^2 / E$ as $k_{\perp }^2 / 2 m_e \times 2/a_e$
in order to explicitize the relationship with eq. (\ref{B04}). 
In the  non relativistic limit $k^2_{\perp} <\!\!< m_e^2$, it is indeed
legitimate to consider $k_{\perp }^2 / 2 m_e$ as providing 
the energy levels of the ``detector states'',
see \cite{Myr}\cite{Stephens} and point b) above.
The manifest similarities between eq. (\ref{B05}) and both 
eq. (\ref{kperp}) and eq. (\ref{B04}) strongly invite
to inquire into their
dynamical origin, if any.

A first indication that there is a deep 
relation is furnished by an analysis
of the euclidean instanton\footnote{Remarkably,
this analysis can be straightforwardly extended to black hole
pair production and their subsequent thermal effects.
Furthermore these relations between euclidean gravity
and thermal phenomena shed a new light on the thermodynamical approach
to gravity\cite{jac} presented by T. Jacobson, see Section 7.}
 associated with the Schwinger process.
This instanton is obtained by considering the classical dynamics
of a relativistic particle of mass $M$ and charge $e$ in an
electric field $E_0$. The classical (Lorentzian) trajectories 
have uniform acceleration either to the right (corresponding to
particles) or to the left (for antiparticles) and the
 euclidean orbits are closed trajectories, as in a magnetic field.
The euclidean action for completing an orbit is
\ba
S_{euclid} = \pi M^2 /E
\label{Seuc}
\ea
Contact with the second quantized framework is made by the
fact that the probability of creating pairs
scales like 
$e^{-S_{euclid}}$, see eq. (\ref{01}).

What really concerns us it the amount of 
euclidean proper time necessary to complete this orbit. It is given by
the Hamilton Jacobi relation
 \be
\tau_{euclid} = \partial_M S_{euclid} = \partial_M 
\left( { \pi M^2 \over E} \right)  = { 2\pi \over a} = T_U^{-1}
\label{prop}
\ee
It equals the inverse Unruh temperature.
At this point, it should be recalled that the quantum processes
induced by the uniform acceleration and leading to 
eq. (\ref{B04}) are {\it all} governed
by lapses of proper time $\tau$.
By using eq. (\ref{prop}), eq. (\ref{B05}) can be written as
\be
{R_{k_{\perp } \rightarrow 0 } \over
R_{0 \rightarrow k_{ \perp } }} =
e^{- {k_{\perp}^2 \over 2m_e} \partial_m S_{euclid}}
\label{infin}
\ee
This strongly suggests that the Unruh process can be obtained
from 
a first order comparison of two neighboring Schwinger processes,
in a manner similar that canonical distributions characterized by a temperature
are obtained from a first order change applied to micro-canonical
distributions characterized by energy only. The validity of this comparison with
thermodynamics will be 
confirmed upon considering black hole
pair production. It will then be clear that the euclidean action behaves
{\it as} an entropy in delivering the Unruh temperature, see eq. (\ref{prop}),
therefore it behaves like the Bekenstein entropy in the latter's
 determination of Hawking temperature.

In this paper we shall prove these
interpretations are correct and we shall provide 
the physical rationale behind them.  To this end, we shall 
use a simple model and proceed in several steps. These are
presented in the next Section.

\section{The model and the strategy}

Instead of working with the transverse momentum, see 
eqs. (\ref{kperp}, \ref{B05}), to establish the relations between the Schwinger
and Unruh effects, we shall use a two-dimensional model\cite{rec}
composed of two charged fields,
$\psi_M$ and $\psi_m$ and the scalar massless field $\phi$. 
The Hamiltonian which governs the transition amplitudes  is
\ba
H_{\phi\psi} = \tilde g \int\! dx \left[ \psi_{M} \psi^\dagger_{m } \phi + h.c. \right]
 \label{B04B}\ea
where $\tilde g$ is a coupling
constant.
We have chosen that model because the expressions 
for the transition amplitudes are
considerably simpler than in the four dimensional case, 
see \cite{Niki2}\cite{Niki}. Thus,
 they display more clearly the seeked relationships.

In Section 5, we shall compute ``exactly''
the first order (in $ \tilde g$) transition amplitudes.
By exactly we mean to all orders in $\Delta m/ M$ and $a /M$ 
thereby taking  into account recoil effects (first quantized effects)
and vacuum instability (a second quantized effect).
In \cite{BPS} both recoils and pair creation effects were neglected,
and in \cite{rec}\cite{rez} 
only recoils effects were taken into account.

>From the properties of these amplitudes under {\it crossing symmetry}, we 
prove that the ratio of the transition rate from the ground 
to the excited state ($m \to M$) to
the inverse transition rate ($M \to m$) is {\it exactly} given by
\be
{R_{m \to M}\over R_{M \to m}}= e^{- \pi (M^2 -m^2)/E} = 
 e^{-2\pi (M-m)/\bar a}
\label{2p}
\ee
In the second equality, as in eq. (\ref{B05}), we have rewritten the exponent
in order to make contact with the Unruh expression
controlled by an acceleration and an energy gap, see eq. (\ref{B04}).
The only difference 
is that the unique acceleration 
is replaced by the mean acceleration $\bar a = (M + m) / 2E$.
This should cause no surprise since the two levels $M$ and $m$
experience different accelerations. In fact the
Unruh formula, eq. (\ref{B04}), should always be considered as an 
approximate expression valid when it is legitimate to deal with
a single acceleration. Strictly speaking, this requires that 
the limit $\Delta m/M \to 0$, $a/ M \to 0$ be taken.
More physically, it requires that the mass scales be well separated,
i.e. $M-m <\!\!< M$. In that case,
the concept of a single\footnote{
These features arise whenever one enlarges the dynamics 
so as to abandon the background 
field approximation wherein it is postulated that all processes can be
described by a quantum system coupled to a single 
trajectory. Indeed, upon studying {\it afterwards} the semi classical regime
to determine how the  background 
field approximation re-emerges, one explicitizes the conditions
 which must prevail to validate that approximation. 
The interested reader will consult \cite{wdwgfpt} where this
approach is applied to quantum cosmology to determine the validity
of the semi-classical Einstein equations.}
acceleration is meaningful, and therefore
the concept of temperature as well.

However, even when the condition $M-m <\!\!< m$ is not satisfied, 
{\it the two level ion reaches equilibrium with the
population ratio of its excited and ground state given by 
the Schwinger mechanism}. Indeed, 
the ratio of the probabilities $P_M$ and $P_m$ to find a 
given particle produced from vacuum in the 
$M$ or the $m$ state is equal to the ratio of the 
mean numbers $N_M$ and $N_m$ 
of produced quanta of masses $M$ and $m$, see  eqs. (\ref{N}, \ref{kperp}),
and therefore given by
\be
{ P_M \over P_m } = e^{- \pi (M^2 -m^2)/E} = \left(
{N_M \over N_m } \right) 
\label{equil34}
\ee
The equality of the ratios of the transition rates, eq. (\ref{2p}), and 
of the probabilities, eq. (\ref{equil34}), 
proves that there is a consistency between the  
Schwinger mechanism and the
extended-Unruh-effect defined by {\it keeping} the finite corrections
in $a/M$ and in $\Delta m/ M$ into account. 
When the mass ratio $\Delta m/ M$ is negligible, one fully recovers the
thermal equilibrium governed by a temperature, as in the original
Unruh description.
When $\Delta m/ M$ is not negligible, even though
it is illegitimate to deal with a single accelerated
trajectory, the
equilibrium distribution 
still exists and still coincides with the Schwinger
distribution. In this sense, pairs are born in equilibrium\cite{BPS}.

In Section 6, we determine the origin of the equality
between eqs. (\ref{2p}) and (\ref{equil34}). We show that this equality is 
{\it dictated} by the analytical 
properties of the amplitudes governing pair creation 
under crossing symmetry and {\it CPT}.

To understand this result, it is necessary to first consider the amplitude
for a particle
to propagate from $t=-\infty$ to $t=+\infty$ (figure 1a). By {\it crossing
symmetry}, one replaces the incoming particle by an outgoing anti-particle,
hence one obtains the amplitude for pair creation
(figure 1b). The currents associated with these two propagations are 
related by the factor $e^{- \pi M^2 /E}$.
\begin{figure}
\begin{picture}(200,200)(0,0)
\mbox{\epsfxsize=100mm \epsfbox{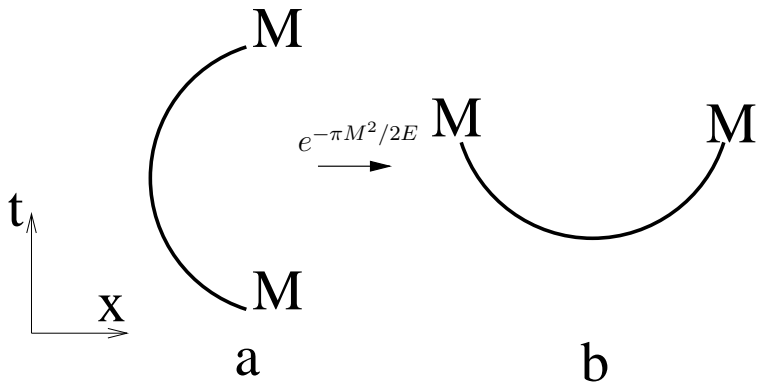}}
\put(-175,90){$e^{-\pi M^2/2E}$}
\end{picture}
\caption{Feynman diagrams representing  (a) 
a particle of mass $M$ and charge $e$ propagating from $t=-\infty$ to
  $t=+\infty$ in the electric field $E_0$, and (b)
a particle--antiparticle pair 
 creation in the electric field. The particles are
  represented by curved lines because they are accelerated by the
  electric field. These diagrams are oriented both in space and
  time: the left-right symmetric of (a) would represent an
  antiparticle accelerated in the opposite direction, the up-down
  symmetric of (b) would represent particle--antiparticle
  annihilation. 
Amplitudes (a) and (b) are related by
  level crossing. The ratio of their norms is given by  the
  Schwinger factor 
  $e^{-\pi M^2/2E}$, see eq. (
) in the text for the 
precise mathematical definition.}
\end{figure}
\begin{figure}
\begin{picture}(300,300)(0,0)
\mbox{\epsfxsize=150mm \epsfbox{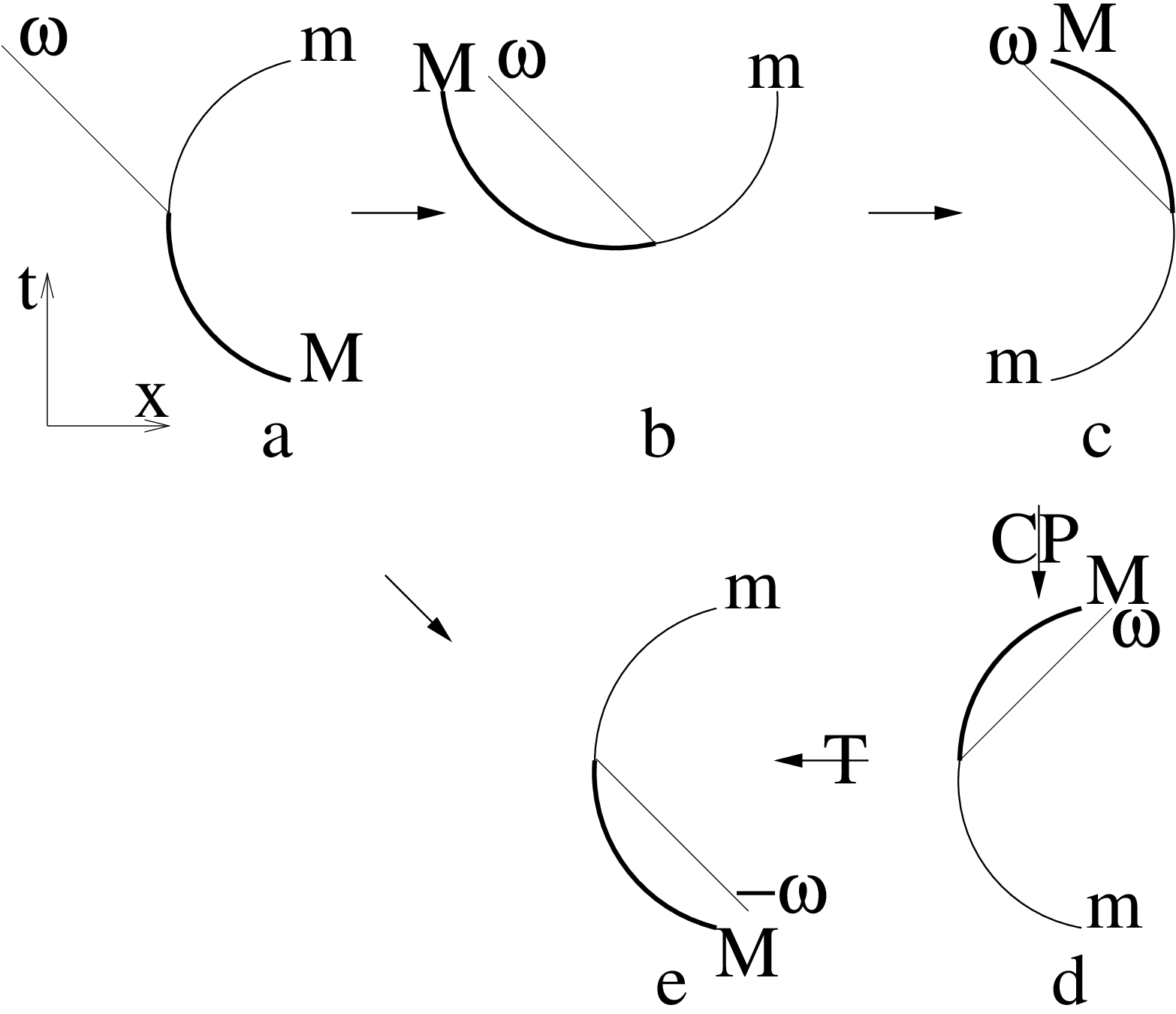}}
\put(-110,295){$e^{+\pi m^2/2E}$}
\put(-310,295){$e^{-\pi M^2/2E}$}
\put(-330,145){$\omega \to e^{i \pi} \omega$}
\end{picture}
\caption{Feynman diagrams representing (a) an accelerated detector
  which deexcites and emits a $\phi$ quantum of energy $\omega$, (b)
 pair creation of
  a pair of detectors and the emission of a $\phi$ quantum, (c)
  spontaneous excitation of an anti--detector, (d) spontaneous
  excitation of a detector, and (e) deexcitation of a detector
  accompanied by the absorption of a $\phi$ quantum. 
The conventions are the same as in figure 1. A thick curved line designates
an excited detector of mass M, a thin curved line a deexcited detector of
mass m, and a straight line a $\phi$ quantum. The orientation of
the straight lines (45 degrees to the right or left) corresponds to
the momentum of the light quantum being $k_x = - \omega$ or $k_x = +\omega$.
Amplitudes (a)
  and (b),  and amplitudes (b) and (c) are related by level
  crossing. Upon passing from one to the other they acquire the
  Schwinger factor $e^{-\pi M^2/ 2E}$ and $e^{\pi m^2/ 2E}$ respectively. Diagrams
  (c) and (d) are related by CP and diagram (d) and (e) by T symmetry.
  Alternatively one can pass directly from (a) to (e) by taking
  $\omega$ to $e^{i\pi} \omega$. 
These relations shall be proven in Section 5 and 6.}
\end{figure}

Now consider radiative processes involving a $\phi$ quantum, 
to first order in 
$\tilde g$.
The amplitude (a) of figure 2 represents the deexcitation of
a detector of mass $M$ accompanied by the emission of a massless quantum of
energy $\omega$. 
The norm squared of this amplitude determines the rate $R_{M\to m}$ of
eq. (\ref{2p}). On the other hand the rate $R_{m \to M}$ of
spontaneous excitation of the detector is determined by the norm squared of
the amplitude (d) of figure 2. Our aim is to understand how
these two amplitudes are related.

First note that by using $T$ symmetry, amplitude (d) is equal up to complex
conjugation to the amplitude (e) for a detector in the excited state to
absorb a light quanta and deexcite. Then note that amplitudes (a) and (e)
are related by crossing symmetry for the light quantum, i.e. by taking
$\omega \to e^{i\pi} \omega$. This latter relation
will be exploited in Section 5. However it does not help to understand
 how eq. (\ref{2p}) relating the rate $R_{m\to M}$ to $R_{M\to m}$ is 
connected to the Schwinger process, eq. (\ref{equil34}), since it involves
crossing symmetry applied to the photon only.

To obtain this understanding, we shall use instead the succession 
a$\to$b$\to$c$\to$d which proceeds through crossing symmetry
applied {\it twice} to the Schwinger process.
First we use crossing symmetry for the 
``excited'' detector of mass $M$ so as to
relate (a) to the amplitude to 
create from vacuum a deexcited detector, an excited
anti--detector, and a massless quantum of energy $\omega$ (figure 2b).
The ratio of the norms of amplitudes (b) and (a) is given by the
Schwinger factor $e^{-\pi M^2/2E}$ exactly as in the case considered
in figure 1 above. Secondly, by using again crossing symmetry 
applied to the deexcited
detector of mass $m$, one obtains
the amplitude (c) for a deexcited anti-detector to get excited and to emit
a light quantum. The ratio of the norms of 
amplitudes (b) and (c) is given by the
Schwinger factor $e^{\pi m^2/2E}$. In this case it is the mass $m$ rather
than the mass $M$ which comes up in the exponential weight.
Thirdly, we use the {\it CP} symmetry to map
particles into anti--particles while leaving the electric field $E_0$
unchanged. Thus {\it CP} maps the amplitude (c) onto the sought 
for amplitude (d) for 
a deexcited detector to get excited and emit a quantum $\omega$.
The equality of eqs. (\ref{2p}) and (\ref{equil34}) and the connection between
the (extended) Unruh effect and the Schwinger process is
thus explained by this succession. 

The motivation for our emphasis on {\it CPT} and crossing symmetry 
is that these  analytical properties should hold irrespectively of the
specific model under examination. 
As an illustration of this universality, 
in Section 7, we consider 
pair creation of charged black holes in a constant electric 
(or magnetic) field\cite{(11)}\cite{(1)}\cite{(2)} and the subsequent
emission of quanta through Unruh effect as well as through 
Hawking process\cite{Hawk}. 
We show that there
is once more a complete thermodynamic consistency between 
the production of the black hole pairs and both
of these radiative effects. This thermodynamic consistency
illustrates that the euclidean action eq. (\ref{Seuc}) acts indeed as 
an entropy in delivering the Unruh temperature, see eq. (\ref{prop}),
since it is given in terms of a quarter of a change in area 
and occurs in amplitudes added to the Bekenstein entropy.


\section{The Schwinger effect}

We recall in this section the essential steps necessary to obtain the Schwinger
effect. The reader unfamiliar with pair creation in an electric field may consult
refs. \cite{Niki2}\cite{Niki}\cite{BPS}\cite{GO}.
What differs in our presentation is the emphasis put on the use
of {\it crossing symmetry} in defining and obtaining the Bogoliubov coefficients.
The reason for this emphasis is that crossing symmetry will play a crucial 
role in Sections 5 and 6.

We consider a massive charged scalar field $\psi _M$ in a constant 
electric field $E_0$. In the the homogeneous gauge ($A_t=0$, $A_z= E_0 t$)
$\psi _M$  obeys the Klein Gordon equation
\be
\left [ \partial_t^2 - (\partial_z  - i E t)^2 - \partial_y^2 -
  \partial_x^2 +M^2 \right] \psi_M =0
\label{KGE}
\ee
where $E=e E_0$.
In this gauge 3-momentum is conserved. The transverse momentum 
squared acts like a shift of the mass squared, see
eq. (\ref{N}). From now on however, for reasons of simplicity,
we shall take it to vanish and work in 1+1 dimensions. 
Since the longitudinal momentum $p$ is also conserved, $\psi _M$
can be decomposed as a sum of $e^{i p z}\chi_{p,M}(t)$ where
$\chi_{p,M}(t)$ obeys 
\begin{equation}
\left[ \partial_t^2 + (p-Et)^2 + M^2  \right] \chi_{p,M}(t) = 0
\label{KGE2}
\end{equation}
There are two independent solutions of this equation and their asymptotic
behavior must be used to identify which linear superpositions describe,
{\it in} and {\it out}, particle and anti-particle states. 
Indeed, because of the time dependence of the frequency,
{\it in} particle modes, i.e. solutions of (\ref{KGE2}) carrying
unit positive current for $t\to -\infty$, will be a superposition of
{\it out} particle and antiparticle modes for $t\to \infty$:
\be
\chi^{in}_{M,p} = \alpha_M \chi^{out}_{M,p} + \beta_M 
\bar\chi^{{out}\  *}_{M,-p}
\label{BOG}
\ee
Current conservation requires
\be
\vert \alpha_M \vert ^2 - \vert \beta_M \vert^2 =1
\ee
To obtain the asymptotic behaviors of the various modes,
it is useful to use the following integral
representation, see \cite{PB1}.
For instance, in-modes are given by
\be
\chi^{in}_{M, p}(t)=\alpha_M \int^{\infty}_0 
{du \over \sqrt{8\pi^2}} (u)^{{-iM^2 \over 2E}-{1 \over 2}}
 e^{iE \left[ u^2/4 - (t-p/E)u +  (t-p/E)^2 /2\right]}
\label{App1}
\ee
where the normalization factor is shown to be the 
coefficient $\alpha_M$.
One easily obtains the
Bogoliubov coefficient $\beta_{M}$ from this integral
representation because,
for large $\vert t \vert$, i.e. $\vert t \vert >> M/E$, the integral
receives all its contribution from saddle points 
at $u \to \infty$ and from the region $u \to 0$. One finds that 
the saddle point at $u \to \infty$ 
describes the outgoing particle wave carrying for $t \to \infty$ a current
$\vert \alpha_M\vert^2$.
Instead
the contribution from $u \to 0$ describes, 
for $t\to -\infty$ the incoming branch carrying unit positive current, and
for $t\to + \infty$ the antiparticle branch carrying negative current,
see \cite{PB1} for more details.
$\beta_{M}$ is the ratio of these latter contributions
at small $u$. To 
evaluate this ratio, it is legitimate to neglect
the term in $u^2$ in the exponential and one is left with the 
integral representation of  $\Gamma$ functions:
\be
\beta_{M} =
{\int^{\infty}_0 {du }\ u^{{-iM^2 \over 2E}-{1 \over 2}}\
 e^{iE(-\vert t \vert u + t^2 /2)} \over 
\int^{\infty}_0 {du }\ u^{{-iM^2 \over 2E}-{1 \over 2}}\
 e^{iE(\vert t \vert u +  t^2 /2)} } = -i e^{-\pi M^2 /2E} 
\label{rapsch}
\ee
Note how it is the sign of the exponent of $e^{\pm iE\vert t \vert u}$ which
governs
the ratio of these integrals. By sending $t$ to $e^{i \pi} t$ in the
lower integral, the contribution of
the incoming particle is replaced by the one of the outgoing anti-particle.
This is what we designate by crossing symmetry, see figure 1.
In section 6, we shall see 
 that it is this continuation used twice which implies 
the equality of eqs. (\ref{2p}) and (\ref{equil34}).

The corresponding {\it out}-mode with asymptotic unit
final current directed towards $z=+\infty$
 is obtained by replacing $t$
by $-t$, $p$ by $-p$ and by complex conjugation. 
Thus, 
its integral representation is
\be
\label{App2}
\chi^{out}_{M, p}(t)=
\chi_{M, - p}^{{in} *}(-t)=
\alpha_M \int^{\infty}_0 {du \over \sqrt{8\pi^2}} (u)^{{iM^2 \over 2E}-{1 \over 2}}
 e^{-iE \left[ u^2/4  + (t - p/E)u +  (t - p/E)^2/2 \right]   }
\ee
One also shows that
the anti-particle {\it in}- and {\it out}-modes are given by
\ba
\bar\chi^{{in}}_{M, -p}(t) &=& \chi^{in}_{M, p}(t)\nonumber\\
\bar\chi^{{out}}_{M, -p}(t) &=& \chi^{out}_{M, p}(t)
\ea

In the second quantized framework, the field operator
$\psi_M$ should be decomposed  both
in terms of the {\it in}-modes and {\it out}-modes
\ba
\psi_M &=& \int_{-\infty}^{+\infty}
\! dp \ e^{ipz} \left[ b^{in}_{M,p} \ \chi^{in}_{M,p} + 
 c^{\dagger in}_{M,-p} \ \bar\chi^{{in} *}_{M,-p} \right]
\nonumber\\
&=&
\int_{-\infty}^{+\infty} dp \ e^{ipz} \left[ b^{out}_{M,p} \ \chi^{out}_{M,p} + 
 c^{\dagger out}_{M,-p} \ \bar\chi^{{out} *}_{M,-p} \right]
\label{antipaa}
\ea
Whereupon one obtains the {\it in}-vacuum and the {\it out}-vacuum, 
which are annihilated by the corresponding destruction operators:
\ba
b^{in}_{M,p} \vert 0,in\rangle_M = c^{in}_{M,p} \vert 0,in\rangle_M
&=& 0
\nonumber\\
b^{out}_{M,p} \vert 0,out\rangle_M = c^{out}_{M,p} \vert 0,out\rangle_M 
&=& 0
\ea From the Bogoliubov transformation, eq. (\ref{BOG}), one obtains the mean
number of produced pairs of momentum $p$
\be
N_M=
\ _M\langle 0,in\vert b^{\dagger out}_{M,p} b^{out}_{M,p} \vert 0,in\rangle_M 
= \vert \beta_M \vert^2 = e^{-\pi M^2 /E}
\ee
One can also express the in-vacuum in term of its out-particle content
\be
 \vert 0,in\rangle_M = {Z_M} \prod_p 
e^{ - {\beta_M \over \alpha_M}
 b^{\dagger out}_{M,p}   c^{\dagger out}_{M,p} }  \vert 0,out\rangle_M 
\ee
where $Z_M$ is the amplitude not to produce pairs. Its norm square is
\be
\vert Z_M \vert^2 = \vert \ _M \scal{0,  out}{0, in}_{M} \vert^2 
= \prod_p \vert {1 \over \alpha_{M} }
\vert ^2 =  e^{-\sum_p 
\ln ( 1 + e^{-\pi M^2/E} ) }
\label{Schw}
\ee
One recovers the Schwinger result, eq. (\ref{01}) by noting that 
$\sum_p= ELT /2\pi$ 
when the electric field is turned on
in a box of size $L$ during a time $T$ (if $T, L >> E^{-1/2}$), see \cite{GO}.

Finally we note that the amplitudes represented in figure 1 are
\ba
\mbox{ fig. 1a}\ \ \ &\leftrightarrow&
\ _M \langle 0, out \vert b^{out}_{M,p} b^{\dagger in}_{M,p} \vert
0,in \rangle_M = Z_M / \alpha_M \nonumber \\
\mbox{fig. 1b} \ \ \ &\leftrightarrow&
\ _M \langle 0, out \vert b^{out}_{M,p} c^{out}_{M,-p} \vert
0,in \rangle_M = -Z_M \beta_M / \alpha_M
\label{fig1}
\ea

\section{The Unruh effect}

We recall the essentials of the Unruh effect for a two level atom, in
the 1+1 dimensional case. The reader may also wish to consult
\cite{Unr}\cite{GO}. We shall again put emphasis on the use of 
{\it crossing symmetry} which allows, in this case, to determine the transition
amplitude of the opposite channel in terms of an analytical continuation
in the energy of the photon applied to the amplitude of the
direct channel. The orientation of the continuation is such
that the stability of the vacuum state is guaranteed.

The trajectory followed by the uniformly accelerated detector is
\ba
t &=& a^{-1} sinh a \tau \nonumber \\
z &=& a^{-1} cosh a \tau
\ea
the detector is coupled to a massless field $\phi$ through the
interaction 
Hamiltonian 
\be
\int dzdt\ H_{int}= g \int d\tau \left[ e^{-i \Delta m \tau} \phi
(t(\tau),z(\tau)) \vert + \rangle
\langle - \vert \ + \ h.c.
\right]
\ee
where $\vert + \rangle $ and $\vert - \rangle$ are the excited and
ground states of the detector, $\Delta m$  is the energy gap between
the two states and $g$ is a coupling constant.

The second quantized field $\phi$ obeys the massless Klein Gordon equation in
1+1 dimensions
\be
\left [ \partial_t^2 -\partial_z^2 \right] \phi =0
\ee
The complete set of solutions with positive Minkowski frequency $\om$
are
\be
\varphi_{k_\omega} = { e^{-i\omega t}e^{i k_\omega z} \over
\sqrt{4 \pi \omega}}\quad\quad \omega = \vert k_\omega \vert \ ,\ 
-\infty < k_\omega < + \infty
\ee
$\phi$ can therefore be decomposed as
\be 
\phi = \int_{-\infty}^{+\infty} d k_\omega \left[ a_{k_\omega}
\varphi_{k_\omega} +\ h.c. \right]
\ee
The Minkowski vacuum is annihilated by all $a_{k_\omega}$ operators
\be
a_{k_\omega} \vert 0 \rangle =0
\ee

To first order in $g$ the amplitude for the detector to deexcite and
emit a right moving Minkowski quantum (i.e. $k_\omega = \omega$) is
\ba
A(\Delta m, \om, a)&=& -i\ \langle  -\vert\langle0\vert a_{k_\omega}  \left[
\int dz dt H_{int}  \right] \vert 0 \rangle \vert + \rangle \nonumber\\
&=&-i ga \int^{\infty}_{-\infty} d\tau
e^{-i\Delta m \tau} { e^{i\om e^{-a\tau} /a }\over \sqrt{4\pi \om}} 
\nonumber\\
&=& -ig 
\int^{\infty}_0 {du\over u}  u^{{i\Delta m}}
 {e^{-iu \om} \over \sqrt{4\pi \om}} \nonumber\\
&=& -ig 
\Gamma(i  \Delta m /a) e^{\pi  \Delta m /2a}
{(\om)^{{i \Delta m /a}} \over \sqrt{4 \pi \om}}
\label{amplit}
\ea
where 
the light like variable $u=t-z$ is related to the proper time $\tau$ by
$au=-e^{-a\tau}$.

Similarly the amplitude for an excited detector 
to {\it absorb} this right moving quantum 
and to get deexcited is
\ba
B(\Delta m, \om, a)&=& -i \ \langle  -\vert\langle0\vert 
\left[ \int dz dt H_{int} \right]  a_{k_\omega}^\dagger \vert 0 \rangle \vert + \rangle
\label{amplit2}
\ea
This amplitude is related to that usually considered in the Unruh
effect, namely the amplitude for a deexcited detector to get
spontaneously excited, by $T$ symmetry, that is complex conjugation.

We shall not computed $B(\Delta m, \om, a)$  directly since it is more
instructive to determine it from the amplitude $A(\Delta m, \om, a)$  by exploiting
their analytical properties under {\it level crossing}, i.e. by taking $\omega \to e^{-i \pi}
\omega$. Indeed, $B(\Delta m, \om, a)$  is given by
\be
B(\Delta m, \om, a)= A(\Delta m, e^{-i \pi}\om, a) \times i
\label{BA}
\ee
Using the fourth line of eq. (\ref{amplit}), one obtains 
\be
B(\Delta m, \om, a)= A(\Delta m, \om, a) \times e^{- \pi \Delta m/a}
\label{BA2}
\ee
Therefore the ratio of the transition rates 
 is 
\be
{ R_{- \to + } \over R_{+ \to - }} =
\vert {B(\Delta m, \om, a) \over A(\Delta m, \om, a) }\vert^2
= e^{- 2 \pi \Delta m / a}
\label{ratiooo}
\ee
since $|B/ A|$ is independent of the energy $\om$ of the photon. 
This is 
exactly what one would have obtained
 in a thermal bath at temperature $T_U = a/2 \pi$, see eq. (\ref{B04}).

\section{The Schwinger mechanism and the Unruh effect}

By using the model of the accelerated two level ion
presented in Section 2,
we shall show to order $\tilde g^2$ that the
ratio of the transition rates $R_{ M \leftrightarrow m}$
to emit a photon starting from the ground state ($m$) or the excited state ($M$)  
satisfies eq. (\ref{2p}) even when the vacuum instability
with respect to pair creation is fully taken into account. 
In the next Section, we shall rederive the same ratio from the sole analytical 
properties of the {\it pair creation amplitudes} under $CPT$ and crossing symmetry. 
It is essential that these latter amplitudes do not vanish (i.e. $\beta_{M} \neq 0$)
in order to determine the transition rates through this second indirect procedure.

We first compute 
the amplitude $ {\cal A}(\Delta m, p, \om)$ (depicted in fig. 2a)
to emit a massless quantum starting from the heavier state $M$.
This amplitude 
corresponds to the amplitude $A(\Delta m, \om, a)$ of eq. (\ref{amplit}).
To first order in $\tilde g$, using the momentum conservation, it
is given by
$$
{\cal A}(\Delta m, p, \om) \delta(p-p'- k_{\om} ) = -i
\ _M \bra{0,  out} \ _m \bra{0,  out}  \bra{0} a_{k_{\om}} b_{m, p'}^{ out}
H_{\psi \phi} b_{M, p}^{\dagger, in} \ \ket{0} \ket{0, in}_{m}\ket{0, in}_{M}
$$
\be
\quad =  -i \tilde g
Z_M Z_m 
\delta(p-p'- k_{\om})
\alpha_M^{-1} \alpha_m^{-1}
\int^\infty_{-\infty} dt \; \chi^{{in} *}_{m, p-k_\om}(t) \; \chi^{out}_{M,p}(t) \; 
{e^{i\om t } \over \sqrt{4 \pi \om} }
\label{ov}
\ee
The overall factor $Z_M Z_m $ is the product of the {\it in-out} overlaps,
see eq. (\ref{Schw}).
It appears because the
scattering process happens in the presence of pair production of charged quanta.

Notice that in the limit $M^2/E \to \infty$ at fixed $M-m= \Delta m$ and $M/E=1/a$, 
the integrand of eq. (\ref{ov}) tends uniformly 
to the WKB expression studied in \cite{rec}.  Therefore, by virtue of the analysis
of that paper, ${\cal A}(\Delta m, p, \om)$ tends to the ``Unruh'' amplitude
$ A(\Delta m, \om, a)$, eq. (\ref{amplit}).

In terms of the integral representations of the $\chi$ modes, see
Section 3, and for $ k_\om = \om$, 
we obtain,
\ba
{ {\cal A}(\Delta m, p, \om)
 \over Z_M Z_m }&=&  -i\tilde g
 \int^{\infty}_0 {du_1\over \sqrt{2\pi}} \int^{\infty}_0 {du_2 \over \sqrt{2\pi}}
\ u_1^{{iM^2 \over 2E}-{1 \over 2}} \ u_2^{{im^2 \over 2E}-{1 \over 2}}\ 
{e^{i\om p/E} \over \sqrt{ 4 \pi \om}}
\nonumber\\
&& \quad \quad \quad \quad \quad
{1 \over 2}
\int_{-\infty}^{\infty} d\tilde t  \ 
e^{i\om  \tilde t} \ 
e^{-iE \left[ u_1^2/4 + \tilde t u_1 + \tilde
t^2/2 + u_2^2/4  - (\tilde t+\om/E) u_2 +  (\tilde
t+\om/E)^2/2 \right]} \quad \quad
\label{Ap}
\ea
where we have defined $\tilde t = t- p/E$.
Performing the Gaussian integration over $\tilde t $ and 
introducing the variable $\delta = E u_1 u_2/2$
 one has
\ba
{ {\cal A}(\Delta m, p, \om) 
 \over  Z_M Z_m  } 
&= & -i \tilde g
\int^{\infty}_0 {du_1\over \sqrt{2\pi}} \int^{\infty}_0 {du_2 \over \sqrt{2\pi}}
u_1^{{iM^2 \over 2E}-{1 \over 2}} u_2^{{im^2 \over 2E}-{1 \over 2}}
{e^{i\om p/E} \over 4 \sqrt{ E \om}} e^{-iE \left[ u_1 u_2/2 -  u_2 \om/E + \om^2 /2E^2 
\right]}\nonumber\\
&= &
-i {\tilde g  \over \sqrt{2E} } e^{i(\om p - \om^2 /2)/ E}
\int^{\infty}_0 {du_2\over u_2}  u_2^{{-i(M^2-m^2)/2E}}
 {e^{iu_2 \om} \over \sqrt{4\pi \om}} \nonumber\\
&& \quad \quad \quad \quad \quad \quad \quad \quad \times
\left[ \int^{\infty}_0  {d \delta \over \sqrt{2\pi} E } \ 
(2\delta /E)^{{iM^2 \over 2E}-{1\over 2}} e^{-{i \delta }} \right] 
\label{App41}
\ea
We postpone the evaluation of this expression since the determination 
of the equilibrium requires to know the ratio of the 
transition rates only, see eq. (\ref{B04}).
Therefore we shall {\it relate} ${\cal{A}}$ to the amplitude of the 
inverse process.
One can either consider the amplitude to emit the same quantum 
starting from the ground state ($m$) (see fig. 2d), or the amplitude to {\it absorb} 
this photon starting 
with the excited detector state ($M$) (see fig.2e), since 
one is the time reversal ($T$ symmetry) of the other.
As in Section 4, we consider the second amplitude, denoted by
$ {\cal B}(\Delta m, p, \om)$, since it is {\it given} by
\be
{\cal B}(\Delta m, p, \om)  = {\cal A}(\Delta m, p, e^{i \pi}  \om) \times i
\label{ABAB}
\ee
by virtue of the stability of the vacuum of the photon field, see eq. (\ref{BA}).

>From the dependence in $\om$ in eq. (\ref{App41}), 
exactly like in the third line of eq. (\ref{amplit}), 
one deduces  immediately that
the square of the amplitudes which determines both the rates $R_{M \leftrightarrow m}$
and the equilibrium probabilities $P_{M (m)}$, satisfy
\be
\vert{ {\cal B}(\Delta m, p, \om) \over {\cal A}(\Delta m, p, \om)} \vert^2= 
{ R_{m \to M} \over R_{M \to m}} =
e^{-\pi (M^2-m^2) /E} = {P_M \over P_m}
\label{exactT}
\ee
Therefore we have proven eq. (\ref{2p}) and the fact that
the equilibrium probabilities $P_M$ and $P_m$ defined by these radiative 
processes are
equal to those
defined by the Schwinger process in eq. (\ref{equil34}). 

Furthermore, when
the mass gap $\Delta m$ satisfies $\Delta m <\!\!< M$, it is meaningful to write 
eq. (\ref{exactT}) as
\be
\vert{ {\cal B}(\Delta m, p, \om) \over {\cal A}(\Delta m, p, \om)} \vert^2 =
 e^{-2\pi  \Delta m /\bar a } 
=
\vert { B(\Delta m, \om, \bar a) \over A(\Delta m, \om, \bar a)}
\vert^2
\label{exactT2}
\ee
where $\bar a  = 2E/(M+m) = a_M ( 1 - \Delta m / 2M)^{-1}$.
 Thus, under the above inequality, one fully recovers
the Unruh equilibrium, see eq. (\ref{ratiooo}), governed by a single acceleration
since $a_M = E/M \simeq E/m \simeq \bar a$.

Notice that it is the first time that the concept of acceleration is brought to bear.
It appears through a first order change in the exponential factor. This is
exactly like the recovery of classical trajectories from wave packets. Indeed 
the stationarity condition is a first order change in the energy (or the momentum)
applied to the phase of the wave packet. This emergence of the classical concepts
of acceleration and temperature also bears many similarities with statistical
mechanics since it is also through a first order change in the energy 
that the concept of equilibrium temperature arises from microcanonical
ensembles.  For further discussions see Section 7.

For completeness, we now compute the amplitude ${\cal{A}}$
itself, see eq. (\ref{App41}).
Performing the integrations one gets
\ba
{ {\cal A}(\Delta m, p, \om) 
 \over  Z_M Z_m  } 
&=&-i {\tilde g \over 2E}{ e^{i(\om p - \om^2 /2)/ E}\over \sqrt{4 \pi \om}}
\Gamma(-i(M^2-m^2)/2E) e^{\pi (M^2-m^2)/2 E}
(\om)^{{i(M^2-m^2)/2E}} 
\nonumber\\
&& \quad \quad \quad \quad \quad \quad \quad \quad \quad 
\left[ \Gamma(i{M^2 \over 2E} +{1\over 2}) e^{\pi M^2 /2 E} 
{(E/2)^{-iM^2/2E} \over \sqrt{2 \pi}} \right]
\label{App51}
\ea
In terms of the mean acceleration $\bar a = 2E/(M+m)$,
one finds
\be
{ {\cal A}(\Delta m, p, \om) 
 \over Z_M Z_m }  = 
\left[{\tilde g  \over 2gE}\right] \ A(\Delta m, \om, \bar a) \ e^{i(\om p - \om^2 /2)/ E}
\left[\alpha_M^{-1} 
(E/2)^{-iM^2/2E} \right]
\label{App52}
\ee
where 
 $A(\Delta m, \om, \bar a)$ is the ``Unruh'' amplitude eq. (\ref{amplit})
 for a two-level system to emit a quantum from the heavier state 
when it follows a classical trajectory of uniform acceleration $\bar a$.

>From this expression, one can determine what are the 
physical processes that cannot be described by
the {\it approximate} amplitudes $A(\Delta m, \om, \bar a)$ and
$B(\Delta m, \om, \bar a)$ based on the hypothesis that one can work
with 
a single classical trajectory. An example of such a quantity is given in \cite{rec}.
It is shown that the dynamical additional phase $e^{i(\om p - \om^2 /2)/ E}$ 
leads to decoherence effects which in turn lead to a positive local flux
after a finite amount of proper time. 
This  positive flux cannot be described in the treatment 
based a classical trajectory because there cannot be 
any loss of coherence 
in the over restricted dynamical framework
wherein only the $\phi$ field carries momentum.

\section{{\it CPT} and crossing symmetry}

The aim of this Section is to rederive 
eq. (\ref{exactT})
from the amplitudes governing the vacuum instability under pair creation.
We shall thereby understand why eq. (\ref{2p}) 
and eq. (\ref{equil34}) coincide.

To this end we shall proceed as explained in Section 2, see also figure 2.
We introduce 
two other amplitudes related to the original amplitude 
$ {\cal A}(\Delta m, p, \om)$, eq. (\ref{ov}),
by crossing symmetry. 
The first one is obtained by replacing the incoming particle created by 
$b^{\dagger, in}_{M, p}$
by an outgoing anti-particle destroyed by $c^{out}_{M , -p}$. 
This matrix element, denoted by ${\cal A}^{(2)} (\Delta m, p, \om)$
(see fig. 2b),
gives the amplitude
to create a pair of charged quanta accompanied by the 
emission of the massless $\om$ quantum. 
The second one is defined by replacing  in ${\cal A}^{(2)}$ the outgoing 
particle destroyed by $b^{out}_{m, p'}$ by an incoming anti-particle created by 
$c^{\dagger , in}_{m, -p'}$. 
This is the amplitude, denoted by ${\cal A}^{(3)} (\Delta m, p, \om)$
(see fig. 2c),
for an accelerated anti-particle of initial mass $m$ to emit an $\om$ quantum. 

The second amplitude is given by the following matrix element
$$
{\cal A}^{(2)} (\Delta m, p, \om)\delta(p-p'- k_{\om})  = -i \ _M
\bra{0,  out}  \ _m \bra{0,  out}  \bra{0} a_{k_{\om}} b_{m, p'}^{ out}
c_{M, -p}^{out} \tilde H_{\psi \phi} \ \ket{0} \ket{0, in}_{m}\ket{0, in}_{M}
$$
\be
  \quad\ \quad \quad 
= -i \tilde g  2\pi 
\delta(p-p'- k_{\om}) {Z_M Z_m \over \alpha_M \alpha_m }
\int^\infty_{-\infty} dt \; \chi^{in,*}_{m, p-k_\om}(t) \; \chi^{in,*}_{M,p}(t) \;
{e^{i\om t } \over \sqrt{4 \pi \om} } 
\label{ov22}
\ee
The second equality follows from the fact that in the homogeneous gauge,
the temporal part of the wave function of an anti-particle of momentum $-p$ is 
 equal to the wave function of the particle of momentum $p$, see eq. (\ref{antipaa}).
Then
the only difference with the integrand of eq. (\ref{Ap}) is the sign flip in the factor 
$e^{-iE \tilde t u_1 }$ which arises from the replacement of $\chi^{out}_{M,p}(t)$ by 
$\chi^{in,*}_{M,p}(t)$\footnote{It should be noted that this product of {\it in}-modes
 appears systematically upon
evaluating any amplitude under the double condition ({\it pre-} and {\it post-} 
selection in the Aharonov language) 
that the initial state of the system was the {\it in-}vacuum and that the final
state contains
one specific pair of charged quanta, see \cite{BMPPS}\cite{GO}. },
where the $\chi$ modes are expressed in their integral representation, 
see eqs. (\ref{App1}, \ref{App2}).
Therefore, exactly as in eq. (\ref{rapsch}), one has
\ba
{ {\cal A}^{(2)} (\Delta m, p, \om) \over {\cal A} (\Delta m, p, \om) } = 
 \beta_{M} = -i e^{-\pi M^2 /2E} 
\label{relat2}
\ea
This relation may be understood qualitatively from the fact that 
${\cal A}$ is the decay amplitude $M \to m + \omega$
whereas ${\cal A}^{(2)}$ can be envisaged as 
describing the production of a pair of heavy particles followed by the decay of one of them into
$m+\omega$. Thus 
one expects ${\cal A}^{(2)}
 \simeq{\cal A} \ e^{-\pi M^2 /2E}$.

Similarly, upon considering the amplitude ${\cal A}^{(3)} (\Delta m, p, \om)$ defined by
$$
{\cal A}^{(3)} (\Delta m, p, \om) \delta(p-p'- k_{\om}) = -i \ _M
\bra{0,  out} \ _m \bra{0,  out}  \bra{0} a_{k_{\om}}
c_{M, -p}^{out}  \tilde H_{\psi \phi} c_{m, -p'}^{ \dagger, in} 
\ket{0} \ket{0, in}_{m}\ket{0, in}_{M}
$$
\be
\quad\ 
 =-i \tilde g 2\pi 
\delta(p-p'- k_{\om}) {Z_M Z_m \over \alpha_M \alpha_m }
\int^\infty_{-\infty} dt \ \chi^{out}_{m, p-k_\om}(t) \; \chi^{in,*}_{M,p}(t) \;
{e^{i\om t } \over \sqrt{4 \pi \om} }
\label{ov2}
\ee
one finds that the sign of the linear term in $u_2$ appearing in the Gaussian factor has flipped.
Therefore
\ba
{ {\cal A}^{(2)} (\Delta m, p, \om) \over {\cal A}^{(3)} (\Delta m, p, \om) } &=& \beta_{m}
= -i e^{-\pi m^2 /2E}
\label{ww2}
\ea
As in eq. (\ref{relat2}), this may be understood from the fact that
the amplitude ${\cal A}^{(2)}$ can also be envisaged as describing
the creation of a pair of light particles followed by the spontaneous excitation of one
of them, whereas
 ${\cal A}^{(3)}$  is the spontaneous excitation amplitude. 

Now, by {\it CPT} invariance,  
one obtains 
\be
{\cal A}^{(3)} (\Delta m, p, \om) = {\cal B}(\Delta m, p, \om) 
\ee
 given in eq. (\ref{ABAB}).
Indeed one verifies that the integrand of ${\cal A}^{(3)} (\Delta m, p, \om)$ 
coincides with the one of $ {\cal B}(\Delta m, p, \om)$
under the change of the dummy variable $\tilde t = - \tilde t$. 
Therefore, combining this latter relation 
with eq. (\ref{relat2}) and eq. (\ref{ww2}), one obtains
\ba
{
{\cal B}
(\Delta m, p, \om) \over {\cal A} (\Delta m, p, \om) } &=& { \beta_{M} 
\over \beta_{m}} =  e^{- \pi (M^2 -m^2) /2E}
\label{ww}
\ea
Thus the ratio of the scattering amplitudes is equal to the 
ratio of the Schwinger factors obtained by using crossing
symmetry twice. This is what guaranteed that eqs. (\ref{2p})
and (\ref{equil34}) coincide. QED.

\vskip .4 truecm

In the above calculation we considered the emission  or absorption of
 a right moving quantum $e^{-i \omega (t-z)}$. Had one considered left moving quanta
$e^{-i \omega (t+z)}$, different transition amplitudes would have been obtained since 
parity $P$ is explicitly
broken in our model by the external electric field. However the ratio of amplitudes for
 left and right
moving  is a constant
\ba
{ {\cal A}(right) \over {\cal A}(left) }
= { {\cal A}^{(2)}(right) \over {\cal A}^{(2)}(left) }
= { {\cal A}^{(3)}(right) \over {\cal A}^{(3)}(left) }
= {\alpha_M \over \alpha_m}\  e^{i \omega^2 /2E}
\label{ratio}
\ea
Therefore the ratios eqs.  (\ref{relat2}, \ref{ww2}, \ref{ww}) 
and the equilibrium distribution eq. (\ref{exactT})
are independent of whether left or right moving
particles are emitted.


\section{Pair creation of black holes}

We consider how the above analysis applies to  
pair creation of charged black holes in an external electric field 
which was considered
in refs. \cite{(11)}\cite{(1)}\cite{(2)}.
In the black hole case, the picture is more complicated because 
black holes have themselves an intrinsic temperature, 
the Hawking temperature,
and because the semi-classical description of the production requires
that their Unruh and Hawking temperature 
coincide. We recall that this condition arises from the requirement that the euclidean 
instanton have no conical singularity. For a given electric field $E_0$,
the charge $Q$ of the hole is a function of its mass $M$.
Thus only the probability to produce this one parameter family 
of black holes can be obtained by this semi-classical treatment.

Following \cite{(4')}\cite{(4)}, we express the 
probability to create a pair of 
black holes 
which belong to this family as 
\begin{equation}
P_{M,Q,E_0} = C e^{ (\Delta {\cal A} + {\cal A}_{BH} )/ 4}
\label{27e}
\end{equation}
where ${\cal A}_{BH} (M,Q) $
is the area of the black hole horizon, $\Delta {\cal A} (M, Q, E_0) $ 
is the {\it change} of the area of the 
acceleration horizon induced by the creation of the black hole pair
and $C$ a constant which takes into account the appropriate phase factors,
see eq. (\ref{N}).
As emphasized in  \cite{(4')}\cite{(4)}, $ {\cal A}_{BH} /4$ 
appears  in this
expression 
as furnishing the density of black hole 
states with mass $M$ and charge $Q$ thereby confirming the 
Bekenstein interpretation
of ${\cal A}_{BH}/4= S_{BH}$ as the black hole entropy.

The domain of the one parameter 
family which can be compared with the Schwinger mechanism
is the one in which the black holes are small
compared to the inverse acceleration, i.e. in the point particle limit. 
Then, the change of the area reduces to
\be
{  \Delta {\cal A}  \over 4} = - S_{euclid} =
- \pi M^2 / Q E_0 \label{E1}
\ee
i.e. minus the euclidean action 
 to complete an orbit, eq. (\ref{Seuc}).

In order to make contact with eq. (\ref{prop}) and therefore to show that 
 $S_{euclid}$ acts as an entropy in delivering the 
Unruh temperature, we consider the black hole pair creation probability 
from another point of view: We assume that eq. (\ref{27e}) 
is valid for {\it all} values of $M$ and $Q$ and not only for the black holes 
which belong to the one parameter family.
Then, one can make independent variations of $M$ and $Q$ 
and determine the most probable mass $\underline M$ at fixed $Q$
by extremizing $P_{M,Q,E_0}$ with respect to $M$.  Using eq. (\ref{E1}), one gets
\ba
\partial_M P_{M,Q,E_0}  
&=& P_{M,Q,E_0} \left[ \partial_M \left({ - \pi M^2  \over Q E_0 }\right) + 
\partial_M \left( {{\cal A}_{BH} (M,Q) \over 4}
\right) \right] \nonumber\\ 
0
&=& P_{\underline M,Q,E_0} \left[ - { 1  \over T_U(\underline M, QE_0) }+
  { 1  \over T_{H} (\underline M, Q)}
 \right]
\label{equilTT}
\ea
where we have 
defined the Hawking temperature as usual by
$d {\cal A}_{BH}/4 = dM/ T_H$.  Therefore the equality of the 
Hawking and the Unruh temperature which defined the one-parameter family
is recovered here as determining the most probable mass $\underline M$.
Indeed one verifies that $\underline M$ constitutes a 
maximum of $P_{M,Q,E_0}$ at fixed\footnote{
Together with Ph. Spindel and Cl. Gabriel, we are presently 
investigating more general variations in which $Q$ also varies.
Then the chemical potential induced by the electric field also participates
to the determination of the equilibrium in the usual thermodynamical way.}
  $Q$ and $E_0$. 
In this determination, the euclidean action $ S_{euclid}$ acts 
exactly like the Bekenstein entropy ${\cal A}_{BH}/4$.
This strongly suggests that 
the equality of Hawking and Unruh temperatures
should be understood in the {\it mean} and not
as a {\it necessary} condition that $M$ and $Q$ must satisfy in order to have
black hole production. 
(A similar point of view has been put forward, but some how less explicitly,
in \cite{(4')}.) 

We now turn to the radiative processes which the black
holes undergo as they are accelerated. Indeed the black holes will
both emit   radiation
through the Hawking process and will interact with the Unruh heat
bath of Rindler quanta. 
Because of the thermodynamic nature of the equilibrium
condition eq. (\ref{equilTT}),
one expects that it should be 
preserved when radiative processes are taken into
account. 
We now show that this is indeed the case, and more importantly that
the rates of emission and absorption of photons can be deduced from
the pair creation probability eq. (\ref{27e}). 
To this end, we define the rate 
$R_{M , \nu}^-$ for an accelerated black hole of mass $M$
to emit massless quanta
of boost energy $\nu$ thereby decreasing its mass by $\nu$.
Similarly we define the rate $R_{M - \nu, \nu}^+$ for the inverse
transition, that is the absorption rate of quanta of boost energy $\nu$
by a black hole of mass $M-\nu$, see the 
amplitudes $A$ and $B$ in Section 4.

On the basis of our analysis of accelerated detectors 
presented in Section 5, we conjecture that the amplitudes for 
these processes are related by level crossing and {\it CPT}
to the amplitudes of producing pairs of 
black holes, 
eq. (\ref{27e}) continued outside of the one parameter family. 
If this is correct, the ratio of the transition rates
can be expressed as
\be
{R_{M -\nu , \nu}^+ \over R_{M, \nu}^-} = 
{ P_{M,Q,E_0} \over P_{M-\nu, Q,E_0}}  
= e^{ - \delta S_{euclid}  + \delta {\cal A}_{BH} / 4}
\label{Pratio2}
\ee
for all values of $\nu$ and of $M$, i.e. 
for values no longer restricted to $\nu <\!\!< M$ nor to $M= \underline M$,
(both conditions being required for the semi-classical approximation\cite{Y1}
 to be valid).

The factor $e^{ - \delta S}$ expresses both the conjecture that transition rates
of charged black holes coincide to those of the corresponding
 point like charged particles (same masses, same charge) and the fact that the latter's 
transition rates are governed by $\delta S_{euclid} = S_{euclid}( M, QE_0) 
- S_{euclid}( M- \nu,  QE_0)$, see eq. (\ref{exactT}),
and not by the 
(canonical) expression $2 \pi (M- \nu)/a$ as in the semi-classical
treatment, see eq. (\ref{ratiooo}).
The  factor $e^{\delta {\cal A}_{BH} / 4}$ expresses the conjecture that 
black holes behave like point like particles characterized by a degeneracy
given by $e^{{ \cal A}_{BH} / 4}$. 
This second conjecture has been recently proven 
for  Schwarzshild black
holes in \cite{kk}.
Notice that eq. (\ref{Pratio2}) reduces to the semi-classical calculation 
when $\nu \to 0$ and when $M= \underline M$.
 Indeed, using eq. (\ref{equilTT}), one obtains directly
\be
{R_{\underline M - \nu , \nu}^+ \over R_{\underline M, \nu}^-} = 
{ P_{\underline M,Q,E} \over P_{\underline M-\nu, Q,E}}  
\to_{\nu \to 0} e^{-\nu/ T_U + \nu/ T_H}=1
\label{Pratio3}
\ee
as in \cite{Y1}.

Thus we see that not only 
the area of the black hole horizon  
acts as a {\it reservoir} entropy in delivering the
properties of the radiation (for a 
recent expose 
which makes clear the passage from a microcanonical ensemble to 
canonical considerations
in black hole thermodynamics,
 see Chapter 3.6 in \cite{GO}), but  
more surprisingly, by virtue of eq. (\ref{E1}),
 the (change in) area of the acceleration horizon acts
in the same way.
Therefore the transition rates are directly determined
by the sum of the change of horizon areas:
\begin{equation}
{R_{M -\nu , \nu}^+ \over R_{M, \nu}^-} = \exp{(\delta {\cal A}_{total}/ 4)}
\label{ATOTAL}
\end{equation}

The present analysis 
 sheds new light on the thermodynamical 
approach to gravity recently presented by Jacobson\cite{jac}.
We recall that his approach is based on two main hypothesis, 
namely that changes in area
are linearly related to changes in entropy and that the surface gravity 
is related to the temperature seen by accelerating observers. From 
these hypothesis, he deduced Einstein equations in the
limit of small fluxes.
The present analysis can be conceived as providing 
statistical (microcanonical) foundations to his thermodynamical 
approach, at least for the restricted set of phenomena considered in this paper.
 Indeed, both of his hypothesis are now derived from the 
fact that transition probabilities are given by the change in
horizon area, eq. (\ref{ATOTAL}). 
(Notice that this expression necessitates a choice
of the action 
that governs the {\it transition amplitudes} of gravity, most likely 
that gravity is described by the 
Einstein-Hilbert action). From eq. (\ref{ATOTAL}) one obtains, first, 
that 
the area of the horizon indeed behaves like an entropy in its
determination of transition rates and equilibrium configurations, 
and, secondly, that acceleration and temperature 
are correctly related.

Finally we note that the local 
interactions between the radiation field and
the black holes lead to a {\it decoherence} of the black holes
states. 
To understand this 
decoherence, note that
before the first
photon is emitted,
 one has a strict EPR correlation between the momenta (and the other 
quantum numbers)
of the two black holes since they are produced from vacuum. 
This correlation is however necessarily destroyed by photons
since the interactions among the radiation field and the black holes
are {\it local} in the sense that the inverse acceleration characterizing the mean
wave length of the radiation emitted or absorbed is much
larger than the horizon radius. Thus 
their masses will spread independently around the mean.
However, in spite of this destruction of the initial correlations,  
the equilibrium distribution of the decohered momenta and masses
is identical to the initial distribution when they were still 
exactly correlated, since the radiative processes maintain
the ``pair creation'' equilibrium, see eq. (\ref{Pratio3}).
%

Moreover, this decoherence is just what it is necessary to invalidate the 
conclusions of the analysis of Yi \cite{Y1}\cite{Y2}. 
He argued that 
accelerated black holes no longer emit radiation when
their Hawking and Unruh temperatures 
coincide. His reasoning was based on coherently interfering amplitudes,
a misleading feature arising when one works in a single classical
background, i.e. by neglecting all quantum recoil effects. As 
stressed in \cite{MaPa2} 
and the last remark of Section 5, the amplitudes evaluated in the background
field approximation are approximations 
which neglect the important phase appearing in eq. ({\ref{App52}). 
Taking into account this phase completely modifies the local properties of the emitted
radiation\cite{rec}.


In summary we have shown that there is a thermodynamic consistency between the
Schwinger and Unruh effects. The classical concept of acceleration,
and the thermodynamic concept of temperature, arise 
upon
taking first order changes in  the energy
applied to the exponential factor appearing in transition amplitudes,
see the remark made after eq. (\ref{exactT}). This is a universal
feature. For instance the 
emergence of time in quantum cosmology also results from first order
treatment of exponential factors\cite{wdwgfpt} 
(the analogous treatment of exponential factors  in  statistical mechanics is
also discussed in this paper). In the case of accelerated black holes 
the consistency with
thermodynamics is enlarged once 
the additional fact that the black holes have an
intrinsic entropy is properly taken into account.
This enlarged consistency can probably 
be derived by appealing to the analytical properties of the
amplitudes to produce black holes and to emit Hawking radiation under
crossing symmetry and {\it CPT}, in close analogy to what we proved for
 accelerated particles. This might shed new light on the debate about whether
black hole evolution can be described by a unitary S-matrix\cite{H2}.

Thus the outcome  of our analysis is that upon enlarging the
dynamical framework and going beyond the semi classical approximation, 
apparently unrelated phenomena such as the Unruh effect,
the Schwinger effect, Hawking radiation, are described in one
thermodynamicaly consistent whole.  And the area of causal horizons
seem to play an essential role in bringing about this unified
description. We shall report further on this aspect in a forthcoming publication\cite{NNN}.

\vspace{1.5 truecm}
{\bf Acknowledgments}

R. P. is grateful C. Bouchiat, R. Brout, J. Iliopoulos and T. Jacobson
for useful comments 
and S.M. would like to thank N. 
Itzhaki.


\vskip 1.5 truecm


\begin{thebibliography}{999}

\bibitem{schw1} W. Heisenberg and H. Euler,  Z. Phys. {\bf  98} (1936) 714,

\bibitem{schw} J. Schwinger, Phys. Rev. {\bf 82} (1951) 664

\bibitem{Unr}  W. G. Unruh, Phys. Rev. D {\bf  14} (1976) 870

\bibitem{Fulling} S. A. Fulling, Phys. Rev. D {\bf 7} (1973) 2850

\bibitem{Unru2}W. G. Unruh, Phys. Rev. D {\bf 46} (1992) 3271

\bibitem{AM}J. Audretsch and R. M\"uller, Phys. Rev. D {\bf 49} (1994)
 4056; Phys. Rev D {\bf 49} (1994) 6566;
Phys. Rev A {\bf 50} (1994) 1755

\bibitem{MaPa} S. Massar and R. Parentani, ``From Vacuum Fluctuations
  to radiation'', 
gr-qc/9502024
to appear in Phys. Rev. D {\bf 54} (1996)


\bibitem{Myr} N. P. Myhrvold, Ann. of Phys. {\bf 160} (1985) 102

\bibitem{Stephens} C. R. Stephens, Ann. of Phys. {\bf 193} (1989) 255

\bibitem{BPS} R. Brout, R. Parentani and Ph. Spindel, Nucl. Phys. B {\bf 353 }
(1991) 209

\bibitem{Niki2} A. I. Nikishov, Sov. Phys. JETP {\bf 30} (1970) 660

\bibitem{Niki} A. I. Nikishov, Sov. Phys. JETP {\bf 32} (1971) 690

\bibitem{rec}R. Parentani, Nucl. Phys. B {\bf 454} (1995) 227

\bibitem{(11)} G. W. Gibbons, in ``Fields and Geometry'', Proceedings
  of the 22nd Karpacz Winter School of Theoretical Physics, Karpacz,
  Poland, 1986, edited by A. Jadczyk (World Scientific, Singapore, 1986)

\bibitem{(1)} D. Garfinkle and A. Strominger, Phys. Lett. B {\bf 256} 
(1991) 146

\bibitem{(2)} S. W. Hawking and S. Ross, Phys. Rev. D {\bf 52} (1995) 5865

\bibitem{Hawk} S. W. Hawking, Comm. Math. Phys. {\bf 43} (1975) 199 

\bibitem{PB1} R. Parentani and R. Brout, Nucl. Phys. B {\bf 388} (1992) 474


\bibitem{GO} R. Brout, S. Massar, R. Parentani and Ph. Spindel,
Phys. Rep.  {\bf 260}  (1995) 329

\bibitem{BMPPS} R. Brout, S. Massar, R. Parentani, S. Popescu and
Ph. Spindel,  Phys. Rev. D {\bf 52} (1995) 1119

\bibitem{(4')}
F. Dowker, J. P. Gauntlett, S. B. Giddings and G. T. 
Horowitz, Phys. Rev. D {\bf 50} (1994) 2662

\bibitem{(4)} S. W. Hawking, G.T. Horowitz and S. Ross, Phys. Rev. D 
{\bf 51} (1995) 4302

\bibitem{Y1} P. Yi,
Phys. Rev. Lett {\bf 75} (1995) 382

\bibitem{Y2} P. Yi, ``Quantum Stability of Accelerated Black Holes'',
  preprint CU-TP-690,
hep-th/9505021


\bibitem{MaPa2} S. Massar and R. Parentani, ``Comment on ``Vanishing
  Hawking Radiation form a Uniformly Accelerated Black Hole'', P. Yi,
Phys. Rev. Lett 75 (1995) 382'', TAU 2325-96, LPTENS 96/17

\bibitem{H2} S. W. Hawking, Phys. Rev. D {\bf 14} (1976) 2460

\bibitem{rez} B. Reznik, {\it First Order Corrections to the Unruh Effect},
gr-qc/9511033

\bibitem{wdwgfpt} R. Parentani, LPTENS 96/45 and 96/46, gr-qc/9610044
and gr-qc/9610045


\bibitem{jac} T. Jacobson, Phys. Rev. Lett. {\bf 75} (1995) 1260

\bibitem{kk} E. Keski-Vakkuri and P. Kraus, {\it Microcanonical
    D-branes and Back Reaction}, hep-th/9610045

\bibitem{NNN} S. Massar and R. Parentani, {\it Gravitational Instanton for Black
    Hole Radiation}, manuscript in preparation.

\end{thebibliography}
\end{document}